\documentstyle[12pt]{article}
\input psfig.sty
\begin{document}

\baselineskip=17pt
\pagestyle{plain}
\setcounter{page}{1}

%%%%%%%%%%%%%%% DEFINITIONS %%%%%%%%%%%%%%%%%%%%%%%%%%%%%

\def\ATMP{Adv.\ Theor.\ Math.\ Phys.\ }
\def\JHEP{J.H.E.P.\ }
\def\MPL{Mod.\ Phys.\ Lett.\ }
\def\NP{Nucl.\ Phys.\ }
\def\PL{Phys.\ Lett.\ }
\def\PRe{Phys.\ Rep.\ }
\def\PR{Phys.\ Rev.\ }
\def\PRL{Phys.\ Rev.\ Lett.\ }
\def\be{\begin{equation}}
\def\ee{\end{equation}}
\def\bea{\begin{eqnarray}}
\def\eea{\end{eqnarray}}
\def\htau{{\widehat\tau}}
\def\ttau{{\widetilde\tau}}
\def\ta{{\widetilde\alpha}}
\def\tb{{\widetilde\beta}}
\def\tm{{\widetilde m}}
\def\tq{{\widetilde q}}
\def\tu{{\widetilde u}}
\def\tx{{\widetilde x}}
\def\ty{{\widetilde y}}
\def\bz{{\bf Z}}
\def\vev#1{{\langle{#1}\rangle}}
\def\bs{{$\backslash$}}
\def\drbar{$\overline{\rm DR}$}

%%%%%%%%%%%%%% THE PAPER %%%%%%%%%%%%%%%%%%%%%%%%%%%%%%%

\begin{titlepage}
\begin{flushright}
CLNS-99/1648, NSF-ITP-99-140
\end{flushright}
\vfill

\begin{center}
{\huge Comparing Instanton Contributions with}\\
\vspace{3mm}
{\huge Exact Results in N=2 Supersymmetric}\\
\vspace{3mm}
{\huge Scale Invariant Theories}
\end{center}
\vspace{10mm}

\begin{center}
{\large Philip C. Argyres$^{1,2}$ and Sophie Pelland$^1$}\\ 
\vspace{3mm}
{\it $^1$Newman Laboratory, Cornell University, Ithaca NY 14853}\\
{\it $^2$Institute for Theoretical Physics, University of California, 
Santa Barbara CA 93106}\\
\vspace{3mm}
{\tt argyres@mail.lns.cornell.edu,  sp74@cornell.edu}
\end{center}
\vspace{10mm}

\begin{center}{\large Abstract}\end{center}
\noindent
We discuss the general issues and ambiguities involved in matching the
exact results for the low energy effective action of scale invariant
${\cal N}=2$ supersymmetric QCD to those obtained by instanton
methods.  We resolve the reported disagreements and verify agreement
between an infinite series of coefficients in the low energy effective
actions calculated in the two approaches.  In particular, we show that
the exact low-energy effective couplings for $SU(N)$ for all $N$ with
$2N$ fundamental hypermultiplets agree at a special vacuum on the
Coulomb branch where a large unbroken discrete global symmetry makes
the matching of parameters relatively straightforward.
\vfill
\end{titlepage}
\newpage

\renewcommand{\baselinestretch}{1.1}

\section{Introduction}

Following \cite{sw9407,sw9408}, the exact low energy effective actions
on the Coulomb branches of many ${\cal N}=2$ gauge theories have been
deduced.  The basic method used is typically a kind of analytic
continuation, where the constraints of rigid special K\"ahler
geometry, matching onto various weak coupling limits, and matching
onto the strongly coupled $SU(2)$ Yang-Mills solution of
\cite{sw9407}, combine to determine the exact non-perturbative
contributions to the low energy effective action for ranges of
parameters where the microscopic theory is strongly coupled.

Alternatively, one may calculate these non-perturbative contributions
directly using semi-classical instanton techniques.  The expansion in
instanton number provides a power series expansion of the couplings of
the low energy effective theory around weak coupling which should
match that of the exact solution.  Such a program has been carried out
for scale-invariant $SU(N)$ theories in
\cite{ahsw9607,dkm9607,hs9608,is9609,dkm9612,kms9804} where
disagreements with the exact solutions were reported.

In this paper we will resolve these conflicts, and perform an infinite
series of checks relating the two approaches.  We note that the
purpose of this paper is to show how to systematically compare the
predictions of any two proposed solutions of a given ${\cal N}=2$
theory, but not to discuss how to derive a given solution.  In
particular, an ${\cal N}=2$ QCD solution must satisfy an intricate web
of consistency requirements following from the interrelations of the
given theory with other theories with different numbers of colors or
flavors; this paper is silent about these consistency requirements,
effectively assuming that the solutions under examination have already
met them.

The paper is organized as follows.  In the next section we discuss the
general issues involved in comparing the exact solutions to the
results of instanton calculations.  The central point is the freedom
to make non-perturbative redefinitions of the parameters and vacuum
expectation values that enter into the low energy effective action.
This freedom was pointed out in \cite{as9509,mn9601} in comparing
different expressions for the effective actions of the scale-invariant
${\cal N}=2$ $SU(3)$ theory, and has been used in discussions
comparing the semi-classical and exact solutions, resolving the
above-mentioned disagreements for ${\cal N}=2$ $SU(2)$ QCD, and
partially resolving the disagreements for ${\cal N}=2$ $SU(N)$ QCD
\cite{ahsw9607,hs9608,is9609,dkm9611,s9701}.  We extend and clarify
these treatments by making a systematic analysis of the full set of
allowed non-perturbative redefinitions.

In Section 3 we illustrate the use of this freedom in the ${\cal N}=4$
$SU(2)$ model broken to ${\cal N}=2$ by an adjoint hypermultiplet
mass, to resolve the ``mismatch'' reported in \cite{dkm9612}.  In this
case there are, however, no non-trivial checks of the exact result
using only the one-instanton result.

In Section 4 we turn to the scale invariant ${\cal N}=2$
supersymmetric $SU(N)$ theories with $2N$ fundamental flavors.  We
first apply our general matching requirements to show the equivalence
of the various exact solutions for $SU(N)$ with $2N$ flavors proposed
in \cite{aps9505,ho9505,mn9507}, about which there seems to be some
doubt in the literature.

Then in Section 5 we resolve the conflicts reported in \cite{kms9804}
for these theories.  In this case there are in principle many
non-trivial checks that can be performed for each $N$ even at the
one-instanton level; we verify an infinite sub-series of these checks.
In particular we show that the one-instanton results and exact methods
agree for all $N$ at a special point in the moduli space of vacua.

\section{Comparing curves and instantons}

In comparing two solutions for the low energy effective actions of any
theory, it is necessary to identify the most general map between the
parameters and vacuum expectation values (vevs) entering into the two
solutions consistent with dimensional analysis, global symmetries, and
the meaning of parameters {\it versus} vevs in the low energy theory.
In the supersymmetric case this matching of parameters is also
constrained by various supersymmetric selection rules which in the
${\cal N}=2$ cases discussed here imply the holomorphic dependence of
certain low-energy effective couplings on parameters in the theory.

The low energy effective action on the Coulomb branch of a
scale-invariant ${\cal N}=2$ $SU(N)$ theory---the specific focus of
this paper---depends on a complex gauge coupling constant $q$, complex
bare mass parameters $m$, and complex vevs $u_k$.  Let us start by
reviewing the definitions of these variables.

{\it Gauge coupling.} In scale-invariant theories the complex gauge
coupling $\tau = (\theta/2\pi) + i(4\pi/g^2)$ is an exactly marginal
parameter of the microscopic (high-energy) theory.  Because of the
angularity of the theta angle (invariance of the theory under integer
shifts of $\tau$) it is convenient to parameterize the theory by $q
\equiv e^{2\pi i\tau}$, for which weak coupling corresponds to $q=0$.

{\it Masses.} Other parameters entering into the microscopic
Lagrangian are the complex bare masses of hypermultiplet fields.  In
terms of ${\cal N}=1$ superfields, a hypermultiplet in the fundamental
representation of $SU(N)$ can be described by two chiral multiplets,
$Q$ and $\widetilde Q$ transforming in the ${\bf N}$ and
$\overline{\bf N}$, respectively.  The complex mass parameter $m$
enters as the superpotential coupling
\be\label{massdef}
{\cal L} \supset m \int\!\! d^2\theta \, {\rm tr} Q \widetilde
Q  + \mbox{c.c.} \ .
\ee

{\it Vevs.}  Unlike $q$ and $m$, vevs are not parameters in the
microscopic Lagrangian, but instead are coordinates on the moduli
space of vacua of the low-energy effective action.  On the Coulomb
branch, the gauge group is generically broken to $U(1)^{N-1}$, and the
vacua form an $N-1$ complex-dimensional space coordinatized by the
gauge invariant combinations of the vevs $v_a$, $a=1,\ldots,N$, of the
adjoint scalar field component $v$ of the ${\cal N}=2$ vector
superfield $V$.  Classically (at weak coupling), writing $v$ as a
traceless complex $N\times N$ matrix in the gauge indices, the flat
directions have (up to gauge rotations) diagonal vevs with entries
$v_a$
\be\label{tracev}
\vev{v} = {\rm diag} \{v_1, \ldots, v_N\},
\ee
satisfying the tracelessness condition $\sum_{a=1}^N v_a =0$.
A subgroup of the gauge group acts by permutations on the $v_a$, so a
basis of gauge invariant coordinates (at least near weak coupling) on
the Coulomb branch can be taken to be the symmetric polynomials
\be\label{vevdefn}
u_k= (-)^k \sum_{a_1<\ldots<a_{k+1}} v_{a_1} \cdots v_{a_{k+1}},
\ee
for $k=1,\ldots,N-1$.  Note that by the tracelessness condition $u_0 
\equiv 0$.  (Indices from the beginning of the
alphabet ($a$, $b$, $c$, $d$) run from $1$ to $N$, while those
from the middle ($j$, $k$, $\ell$, $m$) will run from $1$ to $N-1$
throughout this paper.)

The above definitions only pin down what is meant by $q$, $m$, and
$u_k$ at weak coupling, since they only refer to how these parameters
appear in the classical (weak coupling) Lagrangian, or the relation of
the $u_k$'s to vevs of a field computed by the classical Higgs
mechanism.  The essential point to bear in mind in matching to the
exact solutions found by the ``analytic continuation'' methods
pioneered in \cite{sw9407,sw9408} is that the above weak-coupling
properties of the parameters and vevs are all that is used to define
them.  Thus any two quantities, say vevs $u_k$ and $\tu_k$, that at
weak coupling satisfy the defining relation (\ref{vevdefn}) but
differ by higher-order perturbative or non-perturbative contributions,
are equally valid candidates to be ``the'' vevs entering in the exact
solution.  In particular, the parameters and vevs of the exact
solution can and therefore generically will differ by such
higher-order contributions from the parameters and vevs defined in any
specific calculational scheme (such as the \drbar\ scheme \cite{s79}
used in supersymmetric instanton computations \cite{h86,fp9503}).

There are, of course, some restrictions on this ``anything that
can happen, will'' principle.  First, the parameters and vevs must
satisfy selection rules arising from global symmetries (perhaps
explicitly or spontaneously broken) of the theory.  For example, there
is a global $U(1)_R$ symmetry in the scale invariant models that we
are considering in this paper, under which the $v_a$'s and $m$ all have
the same charge.  Thus the $u_k$ have charge $k+1$ under this
spontaneously broken symmetry, and any redefinition of the $u_k$'s
must preserve these charge assignments.  Also, the selection rules
arising from the unbroken global supersymmetry imply that $q$,
$m$, and $u_k$ all must enter the low-energy effective action
holomorphically, and so only holomorphic redefinitions of these
parameters are allowed.  A perhaps less obvious restriction
arises from the distinction between parameters and vevs in 
field theories: since for a given set of parameters there is
a whole moduli space of vacua, one cannot modify the definition
of a parameter ($q$ or $m$) by making it dependent on the vevs
($u_k$), though it is allowed to do the reverse.

Even with these restrictions there is a large family of allowed
redefinitions of the parameters and vevs of the theory.  We can be
fairly explicit about the form of this family.  Consider two
descriptions of the theory, one with parameters $q$, $m$, and vevs
$u_k$, and the other with $\tq$, $\tm$, and $\tu_k$.  

{\it Couplings.}  As $\tq$ is a parameter it can only depend on $q$
and $m$, but since it is dimensionless it cannot depend on $m$.  By
supersymmetry $\tq$ must depend on $q$ holomorphically and by its weak
coupling definition, must be asymptotically proportional to $q$ as $q
\rightarrow 0$.  The general relation between $\tq$ and $q$ is then
\be\label{qredef}
\tq = c_0 q + c_1 q^2 + c_2 q^3 + \cdots
\ee
for some complex numbers $c_n$.  {\em Thus, what is meant by the
complex coupling in the exact low-energy effective actions of ${\cal
N}=2$ gauge theories is only determined up to the infinite series of
constants $c_n$.}  The coefficient $c_0$ has the form of a one-loop
threshold contribution (a shift of $1/g^2$), while the $c_n$ for
$n\ge1$ correspond to non-perturbative (instanton) modifications of
the coupling.

There are two further points worth remarking on in connection with
(\ref{qredef}).  The first has to do with cases where there are more
than one mass parameter, say $m_1$ and $m_2$.  One can then form a
holomorphic dimensionless ratio of these masses, $m_1/m_2$, which one
might think will enter into the matching (\ref{qredef}).  However,
this cannot happen since then there would be a limit in the space of
theories where the weak-coupling equivalence of $q$ and $\tq$ would
fail.  An example to make the point clear is $\tq = q + (m_1/m_2)
q^2$, which for any fixed $m_1/m_2$ is an allowed redefinition of the
coupling, but fails in the corner of parameter space where $m_1/m_2
\to \infty$ as $q^{-2}$ since it violates the weak-coupling $\tq \sim
q$ requirement.

The second point has to do with global identifications, or
S-dualities, on the space of couplings.  The redefinition
(\ref{qredef}) need not preserve any ``nice'' S-duality action on the
couplings, for though covariance under S-duality transformations might
be convenient, it is not a physical requirement for any description of
a theory; the physically meaningful notions are the topology and
complex structure of the space of couplings which are invariant under
analytic reparametrizations \cite{a9706}.

{\it Masses.} The parameter $\tm$ can only depend holomorphically on
$q$ and $m$, and by dimensional considerations and agreement with
its weak-coupling definition we have
\be\label{mredef}
\tm = (1 + b_1 q + b_2 q^2 + \cdots) m .
\ee
If there is more than one mass parameter, then they may mix at
higher orders in $q$ consistent with any global flavor symmetries.

{\it Vevs.} The $\tu_k$ can depend on $u_k$, $q$, and $m$ holomorphically, 
and should preserve $U(1)_R$ charge assignments and their weak-coupling
definitions:
\bea\label{uredef}
\tu_1 &=& (1 + a^{(1;1)}_1 q + \cdots) u_1 
+ (a^{(1;0,0)}_1 q + a^{(1;0,0)}_2 q^2 +\cdots) m^2 \nonumber\\
\tu_2 &=& (1 + a^{(2;2)}_1 q + \cdots) u_2 
+ (a^{(2;1,0)}_1 q + a^{(2;1,0)}_2 q^2 +\cdots) u_1 m \nonumber\\
&& {} + (a^{(2;0,0,0)}_1 q + a^{(2;0,0,0)}_2 q^2 +\cdots) m^3 \nonumber\\
\tu_3 &=& (1 + a^{(3;3)}_1 q + \cdots) u_3 
+ (a^{(3;2,0)}_1 q + a^{(3;2,0)}_2 q^2 + \cdots) u_2 m \nonumber\\
&& {} + (a^{(3;1,1)}_1 q + a^{(3;1,1)}_2 q^2 + \cdots) u_1 u_1 
+ \cdots \nonumber\\
& \vdots & \nonumber\\
\tu_{N-1} &=& (1 + a^{(N-1;N-1)}_1 q + \cdots) u_{N-1} + \cdots 
+ (a^{(N-1;0,\cdots,0)}_1 q + \cdots) m^N
\eea
where the $a^{(k;\{k_r\})}_n$ are arbitrary complex numbers.  Recall
that $m$ has $U(1)_R$ charge $1$ while the $u_k$ have charges $k+1$.
Thus the condition that the general term in $\tu_k$,
\be\label{someqn}
a^{(k;\{k_r\})}_n q^n \prod_r u_{k_r}
\ee
(where we have defined $u_0 \equiv m$), have the correct $U(1)_R$
charge is $\sum_r (k_r+1) = k+1$.  Note that for similar reasons that
ratios of masses could not appear in (\ref{qredef}), ratios of masses
or vevs do not appear in (\ref{uredef}).

Finally, we can remove a small part of this ambiguity by dimensional
analysis.  So far we have not specified any scale in either of our two
descriptions of the theory, so we have the freedom to equate, by {\it
fiat}, any pair of variables of the same dimension (and consistent
with the symmetries) to define how scales in one description of the
theory are related to scales in the other.  In Section 5 we will find
it convenient to use this freedom to set $\tu_{N-1}=u_{N-1}$ in the
last equation of (\ref{uredef})---{\it i.e.}, set $a^{(N-1;\cdots)}_n
= 0$.

The general matching formulas (\ref{qredef}--\ref{uredef}) are the
main result of this section.  Their coefficients represent an inherent
ambiguity in any prediction derived from the exact low-energy
effective action of ${\cal N}=2$ supersymmetric gauge theory.  In the
next two sections we will describe two very simple examples of how the
use of these matching relations resolves some apparent contradictions
or ambiguities reported in the literature.

\section{Comparing curves and instantons in SU(2) with massive adjoint
matter}

In the one-instanton analysis of $SU(2)$ with a massive adjoint
hypermultiplet given in \cite{dkm9612}, a mismatch with the exact
curve of \cite{sw9408} is reported.  The authors of \cite{dkm9612}
determine the one instanton correction to the position in the
$u_1$-plane where a component of the hypermultiplet becomes massless.
(Here $u_1=u$ in the notation of \cite{dkm9612}.)  They find the
singularity occurs at
\be\label{dkmresult}
u_1 = {1\over4} m^2 + {3\over2} q \, m^2 + {\cal O}(q^2) m^2 ,
\ee
while the curve of \cite{sw9408} gives
\be\label{swresult}
\tu_1 = {1\over4} \tm^2 + 6 \tq \, \tm^2 + {\cal O}(\tq^2) \tm^2 .
\ee
Here we have designated all the curve quantities by tildes, to
differentiate them from those entering the semi-classical analysis.
(Note that in (\ref{swresult}) $\tu_1 = u$ in the notation of
\cite{sw9408}, and is {\em not} the $\tu$ variable of that paper!)

The resolution of this discrepancy is obvious in light of our matching
relations (\ref{qredef}--\ref{uredef}): what is being computed is the
relation between the $\tu_1$ variable used in the exact solution
to the $u_1$ variable used in the instanton analysis.
Given that to leading (semi-classical) order the authors of \cite{dkm9612}
matched $q$, $m^2$, and $u_1$ to $\tq$, $\tm^2$, and $\tu_1$, so that
\bea
\tu_1 &=& u_1 + a^{(1;1)}_1 q \, u_1 + a^{(1;0,0)}_1 q \, m^2 
+ {\cal O}(q^2 u_1, q^2 m^2), \nonumber\\
\tm^2 &=& m^2 + b_1 q \, m^2 + {\cal O}(q^2 m^2) ,\nonumber\\
\tq &=& q + {\cal O}(q^2) ,
\eea
for some complex numbers $a^{(1;k_r)}_n$ and $b_n$,
the relations (\ref{dkmresult},\ref{swresult}) imply only that
\be
a^{(1;1)}_1 + 4 a^{(1;0,0)}_1 - b_1 = 18 .
\ee
This gives a restriction on the change of variables needed to
match the curve to the instanton results, but does not affect
any physical predictions.  One needs to compute two-instanton
corrections in this theory before one can make a physical check
of the two approaches, as noted in \cite{dkm9612}.

\section{Comparing different curves for SU(N) with 2N flavors}

The main computation of this paper, in Section 5, will be an example
checking that the physical predictions (after fixing the relations
between variables) of the exact curve and of the one-instanton
computations match.  The theories for which we do this are the
scale-invariant $SU(N)$ theories with $2N$ flavors of massless
fundamental hypermultiplets.

We will use the exact curve in the form
\be\label{curvee}
y^2 = \left(x^N - \sum_{k=1}^{N-1} \tu_k x^{N-k-1} \right)^2 
- \tq \prod_{r=1}^{2N} \left(x+{\tm_r\over\sqrt2}\right).
\ee
As this form of the curve does not quite correspond to any
of the forms \cite{aps9505,ho9505,mn9507} appearing in the
literature, and also since some uncertainty about
the equivalence of these curves has been expressed \cite{kms9804},
in this section we will show that all these forms are equivalent
to (\ref{curvee}).

We start with the form of the curve derived in \cite{aps9505}:
\be\label{apscrv}
y^2 = \left(x^N - \sum_{k=1}^{N-1} u_k x^{N-k-1} \right)^2 
- (1-g^2) \prod_{r=1}^{2N} (x+ (g-1) \mu + m_r).
\ee
Here $\mu \equiv (1/2N) \sum_r m_r$ and $g$ is some function
of the coupling which has the expansion
\be
g(q) = 1 - 2q + {\cal O}(q^2)
\ee
near weak coupling.  It is clear that defining
\bea
\tq &=& 1-g^2 = 4q + {\cal O}(q^2), \nonumber\\
\tm_r/\sqrt2 &=& m_r + (g-1) \mu = m_r -{q\over N} \sum_s m_s 
+ {\cal O}(q^2m_s), \nonumber\\
\tu_k &=& u_k,
\eea
makes (\ref{apscrv}) and (\ref{curvee}) equivalent.  Since these
relations are of the form of the allowed redefinitions of variables we
derived in Section 2, this then shows the physical equivalence of
these curves.\footnote{The factor of $\sqrt2$ difference in the mass
parameters reflects an incorrect factor of $\sqrt2$ in the definition
of hypermultiplet masses in \cite{aps9505}.}

We now turn to the form of the curve described in \cite{ho9505}:
\be\label{hocrv}
\ty^2 = \left(\tx^N - \ell \sum_{k=1}^{N-1} u_k \tx^{N-k-1} 
+ {1\over4} L \sum_{s=0}^N t_s \tx^{N-s} \right)^2 
- L \prod_{r=1}^{2N} (\tx+ \ell m_r),
\ee
where
\bea
L(q) &=& 64q + {\cal O}(q^2), \nonumber\\
\ell(q) &=& 1 + {\cal O}(q), \nonumber\\
t_s(m_r) &=& \sum_{r_1>\cdots>r_s} m_{r_1} \cdots m_{r_s}; 
\qquad t_0 \equiv 1 .
\eea
If we rescale the $u_k$ by $\ell$ and shift by $(L/4) t_{k+1}$,
rescale the masses by $\ell$, and define $\tq = L$, (all of which are
allowed redefinitions by our previous discussion) then (\ref{hocrv})
becomes of the form (\ref{curvee}) except for two minor discrepancies:
in the squared term on the right hand side there is an $\tx^{N-1}$
term with coefficient $(L/4) t_1 \sim q m_r$ and the $\tx^N$ term has
coefficient $1+(L/4) \sim 1 + {\cal O}(q)$.  These discrepancies can
be removed by a shift of the (dummy) $\tx$ variable by $\sim Lm_r$ and
a rescaling of $\ty$ by $1+(L/4)$.  These then require shifts and
rescalings of the coupling, masses, and vevs, but since the shifts are
all ${\cal O}(q)$ and the rescalings are $\sim 1+{\cal O}(q)$, they
are also allowed redefinitions of our variables.  Explicitly, the
change of variables
\bea
\tx &=& x + \alpha, \nonumber\\
\ty &=& {1\over4} (4+L) y, \nonumber\\
\tq &=& {16 L\over (4+L)^2}
\qquad \sim \quad 64 q + {\cal O}(q^2),\nonumber\\
\tm_r/\sqrt2 &=& \ell m_r + \alpha 
\,\qquad \sim \quad m_r + {\cal O}(qm_s), \nonumber\\
\tu_k &=& k{N\choose k+1}\alpha^{k+1} + \sum_{j=1}^k {1\over 4+L}
{N-1-j \choose k-j} (4\ell u_j - L t_{j+1}) \alpha^{k-j} \nonumber\\
&& \qquad\qquad\qquad \sim \quad u_k + {\cal O}(qm_r, qu_j),
\eea
where
\be
\,\ \ \ \alpha \ \equiv \ -{1\over N} {L\over 4+L} t_1 
\quad \sim \quad -{16\over N} q \sum_r m_r 
+ {\cal O}(q^2 m_s),\qquad\qquad\qquad
\ee
takes (\ref{hocrv}) to (\ref{curvee}).

Finally, the form of the scale-invariant 6 flavor $SU(3)$ curve
proposed in \cite{mn9507} is
\be\label{mncrv}
\ty^2 = (\xi x^3 - u_1 x - u_2)^2 - (\xi^2-1) \prod_{r=1}^6
(x - {1\over6}(1-\xi^{-1})\sum_s m_s + m_r),
\ee
where
\be
\xi(q) = 1 + {\cal O}(q) .
\ee
Then, very much as above, if we change variables to
\bea
\ty &=& \xi y , \nonumber\\
\tq &=& 1 - \xi^{-2} , \nonumber\\
\tm_r/\sqrt2 &=& m_r - {1\over6}(1-\xi^{-1})\sum_s m_s , \nonumber\\
\tu_k &=& \xi^{-1} u_k ,
\eea
we recover (\ref{curvee}).

\section{Comparing curves and instantons in SU(N) with 2N flavors}

We now come to the main calculation of this paper, where we resolve
the difficulty, reported in \cite{kms9804}, with matching a
one-instanton calculation to the exact curve for the scale-invariant
$2N$ flavor $SU(N)$ theory.  

The authors of \cite{kms9804} proposed to relate the coupling
parameter $\ttau$ of the curve (\ref{curvee}) to the low energy matrix
of effective couplings $\ttau^{jk}$ at a special vacuum in the moduli
space of the theory, where {\em classically} $\ttau^{jk}=\ttau\,
\tau^{jk}_{cl}$ with
\be\label{tauclass}
\tau^{jk}_{cl} \equiv 1 + \delta^{jk} = 
\pmatrix{ 2 & 1 & 1 & \ldots \cr
1 & 2 & 1 & \ldots \cr 1 & 1 & 2 & \ldots \cr
\vdots & \vdots & \vdots & \ddots \cr} .
\ee
The low-energy matrix of $U(1)^{N-1}$ couplings are the
coefficients of the $U(1)^{N-1}$ gauge kinetic term,
\be
- {1\over16\pi} {\rm Im} \tau^{jk} F^+_j \cdot F^+_k , 
\ee
in the low-energy effective action on the Coulomb
branch.\footnote{Here $F^+_j \equiv F^{\mu\nu}_j -{i\over2}
\epsilon^{\mu\nu\rho\sigma} F_{j,\rho\sigma}$ is the self-dual field
strength of the $j$th $U(1)$ factor.}  The problem is then that for
$SU(N)$ with $N>3$ it can be proven \cite{mn9601,ay9601} that
$\ttau^{jk}$ is not proportional to $\tau^{jk}_{cl}$ at {\em any}
point on the Coulomb branch.

The first point to make towards resolving this difficulty is to note
that $\ttau$ is the microscopic coupling parameter of the theory, and
as such should not be expected in general to satisfy any simple
relation to the matrix of low-energy couplings $\ttau^{jk}$.  The
low-energy coupling is some function of the microscopic coupling
parameter and the vevs.  The instanton methods also calculate these
low-energy coupling functions as an expansion in $q$:
\be\label{tauexp}
\tau^{jk}={\ln q\over2\pi i} \tau^{jk}_{cl} 
+ \sum_{n=0}^\infty q^n \tau^{jk}_n
\ee
where $\tau^{jk}_0$ is the one-loop threshold matching contribution,
while $\tau^{jk}_n$ for $n>0$ are the $n$-instanton contributions;
all the $\tau^{jk}_n$ are functions of the
vevs.  The proper way to compare the results of the two methods is to
compare their predictions for the matrix of low-energy couplings as
functions of the coupling and the vevs modulo the allowed
redefinitions (\ref{qredef}) and (\ref{uredef}) of the variables.
(Note that here we are setting all mass parameters to zero.)
There is also the inherent ambiguity in the definition of the
low-energy couplings themselves by electric-magnetic duality
transformations, which must also be taken into account; we will
encounter examples of this below.

To compare the physical predictions of the exact and instanton
methods, we must relate the parameter and vevs on the two sides. On
the exact side, we use the variables $\tq$ and $\tu_k$, while on the
instanton side, we use $q$ and $u_k$.  (In general, quantities
associated with the exact solutions will have tildes, those in the
instanton calculations, not.)  The technical problem of how to carry
out this matching in practice remains, and might seem difficult in
light of the large number of undetermined coefficients in the matching
relations (\ref{uredef}).  However, we can use the following simple
trick to organize the comparison of the two methods: we expand about a
special direction in the moduli space where the $U(1)_R$ symmetry is
partly restored to a $\bz_N$ symmetry and makes the matching in
(\ref{uredef}) unambiguous.  In particular, along the line $u_1 = u_2
= \ldots = u_{N-2} = 0$ in the Coulomb branch (with $m=0$),
(\ref{uredef}) implies simply
\bea
\tu_k &=& u_k = 0, \qquad k=1,\ldots,N-2 , \nonumber\\
\tu_{N-1} &=& u_{N-1} .
\eea
Here we have used our freedom to set dimensionful scales to equate
$\tu_{N-1}$ and $u_{N-1}$ exactly, without any $q$-dependent factor,
as discussed at the end of Section 2.  Indeed, since we will only
be calculating dimensionless couplings, without any loss of generality
we set
\be\label{norm}
\tu_{N-1} = u_{N-1} = 1
\ee
from now on.  We will refer to this point on the Coulomb branch as the
special vacuum.  It turns out to be more convenient to express the
vevs on the instanton side in terms of the eigenvalues $v_a$,
$a=1,\ldots, N$ satisfying the tracelessness condition $\sum_{a=1}^N
v_a =0$, and related to the $u_k$ by (\ref{vevdefn}).  In these
variables, the special vacuum is at
\be\label{spvac}
v_a |_{sv} = \omega^a,\qquad\mbox{where}
\quad \omega \equiv e^{2\pi i/N} .
\ee

The leading terms in the matching relation for the couplings are
\be\label{qmatch}
\tq = C_0 q + C_1 q^2 + {\cal O}(q^3) ,
\ee
for arbitrary complex constants $C_0$ and $C_1$.  (The $C_0$ coefficient
is computed by a one-loop threshold effect.)

We now compare the curve and instanton predictions by computing
the matrix of low-energy couplings in each approach, expanding about
weak coupling at the special vacuum:
\bea
\tau^{jk} &=& {\ln q \over2\pi i} \tau^{jk}_{cl} + 
\tau^{jk}_0 + q \tau^{jk}_1 + {\cal O}(q^2), \nonumber\\
\ttau^{jk} &=& {\ln\tq \over2\pi i} \ttau^{jk}_{cl} + 
\ttau^{jk}_0 + \tq \ttau^{jk}_1 + {\cal O}(q^2).
\eea
Here we have used the fact that the tree-level Higgs mechanism implies
the classical couplings $\tau \cdot \tau^{jk}_{cl}$ where
$\tau^{jk}_{cl}$ is the constant matrix given in (\ref{tauclass});
the off-diagonal elements appear because of the tracelessness
condition for the $v_a$.  Now equating the two expansions term by
term in $q$ using the matching relation (\ref{qmatch}) implies
\bea\label{taumatch}
\tau_{cl}^{jk} &=& \ttau_{cl}^{jk} ,\nonumber\\
\tau^{jk}_0  &=& {\ln C_0\over2\pi i} \tau^{jk}_{cl}
+ \ttau^{jk}_0  ,\nonumber\\
\tau^{jk}_1 &=&  {C_1\over2\pi iC_0} \tau^{jk}_{cl}
+ C_0 \ttau^{jk}_1 ,
\eea
at the special vacuum. If there exist numbers $C_0$, and $C_1$ that
make these equations true, then we will have shown that, to this order
in the coupling, the two methods agree.\footnote{There is a discrete
ambiguity due to electric-magnetic duality redefinitions in the
low-energy theory, which we have suppressed.  This will be discussed
below.} The rest of this paper will be concerned with computing the
coefficient matrices in (\ref{taumatch}) and solving for the $C_0$ and
$C_1$ matching constants to show that the two methods do indeed agree.

\subsection{Curve computation}

We use for the curve describing the auxiliary Riemann surface $\Sigma$
whose complex structure encodes $\ttau^{jk}$ for the scale-invariant 
$SU(N)$ with $2N$ massless flavors the one (\ref{curvee}) discussed
in Section 4, which we reproduce here (with masses set to zero):
\be\label{curve}
y^2 = \left( x^N-\sum_{k=1}^{N-1} \tu_k x^{N-1-k} \right)^2 - \tq x^{2N}.
\ee
The matrix of low-energy couplings, $\ttau^{jk}$, is defined in terms
of data on $\Sigma$ by
\be
\ttau^{jk} = (A^{-1})^j_\ell B^{\ell k},\qquad\mbox{with}\quad 
A_j^\ell = \oint_{\alpha_j} \omega^{(\ell)} \quad\mbox{and}\quad
B^{\ell k} = \oint_{\beta^k} \omega^{(\ell)},
\ee
where the $\omega^{(\ell)}$ are an arbitrary basis of holomorphic
differentials on $\Sigma$ which can conveniently be taken to be
\be
\omega^{(\ell)} = { x^{\ell-1} dx \over y }, 
\qquad \ell = 1 , \ldots , N-1,
\ee
and where $\alpha_j$ and $\beta^j$ form a canonical homology basis 
on $\Sigma$, which is a basis defined to have intersections 
$\alpha_j \circ \alpha_k =\beta^j \circ \beta^k
=0$, and $\alpha_j \circ \beta^k = \delta_j^k$.

\begin{figure}
\centerline{\psfig{figure=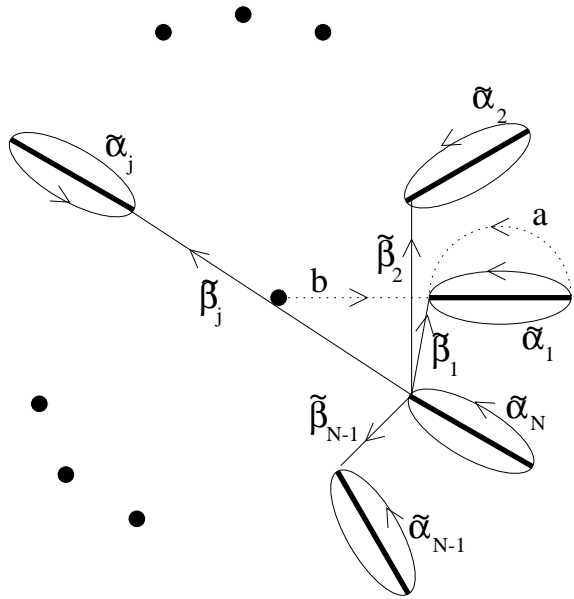,width=3truein}}
{\caption{Representation of one sheet of the two-sheeted covering
of $\Sigma$ given in (\ref{curve}) with a (non-canonical) basis
$\{\ta_j,\tb_k\}$ of cycles in homology.  The dark lines represent
the $N$ cuts on the $x$-plane.  Only half of the $\tb_k$ cycles are
visible, the other half lying on the second sheet.  The dotted
lines represent the $a$ and $b$ contours used in Appendix A.}\label{fig1}}
\end{figure}

To compute the $A^\ell_j$ and $B^{\ell k}$ we must first select a
canonical homology basis. Figure 1 shows a (non-canonical) homology
basis, $\ta_j$ and $\tb^j$, convenient for calculations, in terms of
which a canonical homology basis $\alpha_j$, $\beta^j$ is:
\be
\alpha_j=\ta_j , \qquad
\beta^j=\tb^j+\sum_{k=1}^j\ta_k .
\ee

$A^\ell_j$ and $B^{\ell k}$ are evaluated at the special vacuum in 
Appendix A, yielding
\bea\label{tausvcurve}
\ttau^{jk} &=& {\ln \tq \over 2\pi i} (1+\delta^{jk}) 
+ (1+\delta^{jk}) \left( {1\over2} + {i\over\pi} \ln 4N \right) 
+ {i\over\pi} L^{jk} \nonumber\\
&& {} + \tq {i \over 4\pi N^2} \Biggl[ (1+\delta^{jk})
{1-7 N^2 \over 6} + {2\omega^j \over 
(1-\omega^j)^2} + {2\omega^k \over (1-\omega^k)^2} \nonumber\\
&& \qquad\qquad {} - (1-\delta^{jk}) { 2 \omega^{j+k} \over 
(\omega^j - \omega^k)^2 } \Biggr] ,
\eea
where
\be\label{Ljkdef}
L^{jk} \equiv (1-\delta^{jk}) \ln \left[{\sin(\pi j/N) \sin(\pi k/N) \over 
\sin(\pi |j-k|/N)}\right] + \delta^{jk} \ln \left[ \sin^2(\pi j/N) \right] .
\ee
{}From (\ref{tausvcurve}) we calculate the right sides of
(\ref{taumatch}) to be
\be\label{taumatchcrv}
{\ln C_0\over2\pi i} \tau^{jk}_{cl}+ \ttau^{jk}_0
=  (1+\delta^{jk}) \left[ {1\over2} + {1\over2\pi i} 
\ln \left(C_0\over16N^2\right) \right] + {i\over\pi} L^{jk} ,
\ee
and
\bea\label{taumatchcrvii}
{C_1\over2\pi iC_0} \tau^{jk}_{cl}+C_0 \ttau^{jk}_1 &=& 
{C_0 \over 4\pi iN^2} \Biggl\{ (1+\delta^{jk}) \left[ {N^2-1\over6}
+\left(1+{2C_1\over C_0^2}\right) N^2 \right] \nonumber\\
&& \ {}- {2\omega^j \over (1-\omega^j)^2} 
- {2\omega^k \over (1-\omega^k)^2} + (1-\delta^{jk}) 
{ 2\omega^{j+k} \over (\omega^j - \omega^k)^2 } \Biggr\} .
\eea

\subsection{The electric-magnetic duality ambiguity}

Different choices of canonical homology bases translate into different
$\tau^{jk}$ related by
\be\label{emdual}
\tau \rightarrow (a\tau+b)(c\tau+d)^{-1}
\ee
where ${a\ b\choose c\ d} \in Sp(2N-2,\bz)$ for $(N-1)\times (N-1)$
integer matrices $a$, $b$, $c$, and $d$.  This is an electric-magnetic
duality transformation on the low-energy $U(1)^{N-1}$ couplings, and
is an inherent ambiguity in the meaning of those couplings: any two
$\tau^{jk}$ related by (\ref{emdual}) are physically equivalent.  In
comparing the $\tau^{jk}$'s predicted by the curve and instanton
methods, the possibility that they differ by such an $Sp(2N-2,\bz)$
transformation must be taken into account.

This is only a discrete ambiguity, however, since $Sp(2N-2,\bz)$ is a
discrete group.  In particular, the entries in ${a\ b\choose c\ d}$
are integers and therefore $q$ and $u_k$ independent.  Thus it
suffices to determine this matrix at one vacuum (the special vacuum,
in our case).

In the case at hand, we have chosen our canonical basis of cycles
$\{\alpha_j,\beta_k\}$ so that the leading (classical) term in the
low-energy matches with the semi-classical result: the coefficient of
$(\ln q)/(2\pi i)$ in both cases is $\tau^{jk}_{cl} = 1 +\delta^{jk}$.
We can ask, given this agreement, what electric-magnetic duality
ambiguity remains?  Consider two weak-coupling expansions of the
low-energy $\tau_{ij}$ of the form (\ref{tauexp}) with coefficients
$\tau^{jk}_n$ and $\htau^{jk}_n$.  Then the condition that they be
related by an $Sp(2N-2,\bz)$ transformation of the form (\ref{emdual})
for arbitrary (weak) coupling $q$ implies that $c=0$ and
\bea\label{emshift}
a \tau_{cl} &=& \tau_{cl} d,\nonumber\\
a \htau_0 + b &=& \tau_0 d,\nonumber\\
a \htau_n &=& \tau_n d,\qquad n\ge 1 ,
\eea
where we are treating $\tau$ and $\htau$ as $(N-1)\times (N-1)$
matrices, and matrix multiplication is understood.  The first thing to
note is that $b$, which corresponds to constant shifts of the
$\tau^{jk}$, is determined at the one-loop level; we will determine it
below.  Secondly, the condition that ${a\ b\choose 0\ d} \in
Sp(2N-2,\bz)$ implies that $d = (a^T)^{-1}$, so the only remaining
ambiguity lies in the choice of $a\in SL(N-1,\bz)$ which satisfies
\be\label{emcond}
a \tau_{cl} a^T = \tau_{cl} .
\ee

We can determine the possible $a$ satisfying (\ref{emcond}) as follows.
Think of $a$ as a matrix mapping the lattice of integer $(N-1)$-vectors
to itself.  (\ref{emcond}) implies that $a$ also preserves the ``metric''
$\tau_{cl}$ on this space.  But with this metric, the minimum non-zero
length-squared of an integer vector is $2$, given by vectors of the
form $\pm e_j$ or $e_j-e_k$ where $e_j$ is the vector with a $1$ in the
$j$th position and zeros elsewhere.  So $a$ must map the basis $\{e_i\}$
to another basis made up of $\pm e_j$ and/or $e_j-e_k$.  Up to a permutation
matrix, such a matrix must be of the triangular form
\be
a_\triangle = \pmatrix{
\pm1  & *    & *    &\cdots& *    \cr
0     & \pm1 & *    &\cdots& *    \cr
0     & 0    & \pm1 &\cdots& *    \cr
\vdots&\vdots&\vdots&\ddots&\vdots\cr
0     & 0    & 0    &\cdots& \pm1 \cr}
\ee
where the upper triangular entries are either $0$ or $\pm1$ with at
most one non-zero off-diagonal entry in each column, of opposite
sign to the diagonal entry in that column.  The general form
for $a$ is then
\be
a = P_1 a_\triangle P_2,
\ee
for $P_{1,2}$ arbitrary permutation matrices.

It so happens that the matching at the special vacuum using the
canonical basis of Fig.~1, which we perform below, works for $a=1$, so
we do not have to exercise the option of more complicated 
electric-magnetic duality transformations with $a\neq 1$.  We will
see, however, that we will have to use our freedom to do integer
shifts of $\tau^{jk}$, which correspond to electric-magnetic duality
transformations with $b\neq 0$.

\subsection{One-loop threshold computation}

The one-loop correction is a threshold correction, i.e., a
renormalization scheme dependent difference between the gauge
couplings in the high energy theory and those in the low energy
effective theory.  Such threshold corrections were calculated in
\cite{w80} in a dimensional regularization with minimal subtraction
scheme, and were converted to the \drbar\ scheme in \cite{mv9308},
giving the result
\bea
{1\over g_{ab}^2} - {1\over g_0^2} &=& -{1\over48\pi^2} \Biggl\{
-21 \sum_V {\rm Tr} \left( t_a^V t_b^V \ln {M_V\over\mu} \right)
+ 8 \sum_F {\rm Tr} \left( t_a^F t_b^F \ln {M_F\over\mu} \right)
\nonumber\\
&& \qquad\qquad{} 
+ \sum_S {\rm Tr} \left( t_a^S t_b^S \ln {M_S\over\mu} \right)
\Biggr\}.
\eea
Here $\mu$ is an energy scale, $g_0^2$ is the high energy coupling,
$g^2_{ab}(\mu)$ are the low energy couplings, $M_{V,F,S}$ are mass
matrices, the $t_a$'s are the properly normalized generators of the
low energy unbroken $U(1)$'s, and the sums are over (real) massive
vectors, Dirac fermions, and real scalars.  Since a massive ${\cal
N}=2$ vector multiplet has one massive vector field, one Dirac fermion
field, and one real scalar field, all in the adjoint representation,
while our $2N$ massive hypermultiplets each have a Dirac fermion and
four real scalars, all in the fundamental representation, we get
\be\label{thresh}
{4\pi i\over g_{ab}^2 (\mu)} - {4\pi i\over g_0^2} = {i\over\pi} \left[
{\rm Tr} \left( t_a^V t_b^V \ln \left| {M_V\over\mu} \right| \right)
- 2N{\rm Tr}\left(t_a^H t_b^H \ln \left| {M_H\over\mu} \right| \right) 
\right] .
\ee
Here $t_a^V$, $t_a^H$ are generators of the low energy $U(1)^{N-1}$
gauge group descended from the Cartan generators in the adjoint and
fundamental representations, respectively, of the microscopic $SU(N)$
gauge group.  It is easiest to calculate in the case where the high
energy theory is $U(N)$ broken down to $U(1)^N$ and then decouple the
extra diagonal $U(1)$ factor at the end.  The conventionally
normalized generators are then
\be\label{gens}
(t_a^V)^{bd}_{ce} = {1\over\sqrt2}\delta^b_e \delta^d_c
\left( \delta^b_a - \delta^a_c \right), \qquad
(t_a^H)^b_c = {1\over\sqrt2}\delta^b_c \delta^b_a ,
\ee
where $a,b,c,d,e = 1, \ldots, N$ and we are thinking of the $N$ $U(1)$
generators $t^V_a$ as matrices mapping the $N^2$-dimensional space of
adjoint $bc$ indices to the $N^2$-dimensional space of $de$ indices,
and the $t^H_a$ similarly as $N\times N$ matrices acting on the
$N$-dimensional fundamental representation space.

The vector multiplet mass matrix, by the usual Higgs mechanism, is
read off from the kinetic term for $v$, giving
\be\label{vm}
\mbox{Tr}\left(\left[A^{\mu},\vev{v}\right]^\dagger
\left[A_\mu,\vev{v}\right]\right)=-\left|v_a-v_b\right|^2
\left(A^\mu\right)^a_b\left(A_\mu\right)^b_a ,
\ee
implying that the diagonal $A$'s are massless, and the off-diagonal
$A$'s have mass $|v_a-v_b|$, giving the mass matrix
\be\label{vM}
\left| M_V \right|^{ac}_{bd} = 
\left| v_a - v_b \right|\delta^a_d\delta^c_b.
\ee
The masses for the accompanying massive real scalar and Dirac fermion
in the ${\cal N}=2$ vector multiplet are the same by supersymmetry.
The hypermultiplet masses arise from the superpotential term
\cite{gsw78}
\be\label{hm}
{\cal L} \supset \sqrt{2} \int\! d\theta^2\, {\rm tr}(Qv\widetilde Q)
+ \mbox{c.c.} \ .
\ee
Replacing $v$ by its vev generates a mass term for the hypermultiplets
which, comparing to the bare mass term (\ref{massdef}), implies the
hypermultiplet mass matrix
\be\label{hM}
\left|M_H\right|^a_b = \sqrt{2} \left| v_a\right| \delta^a_b .
\ee
These are the masses for the four real scalars as well as the Dirac
fermion in the massive hypermultiplet.

Plugging (\ref{gens},\ref{vM},\ref{hM}) into (\ref{thresh}) gives
\be\label{threshii}
i {\rm Im} \tau_0^{ab}
= {i\over\pi} \left[
\delta^{ab} \left( \sum_{c\neq a}^N \ln|v_a-v_c| 
- N \ln|\sqrt2 v_a| \right) -
(1-\delta^{ab}) \ln|v_a-v_b| \right] ,
\ee
where we have used $\tau^{ab} \equiv (\theta^{ab}/2\pi) + i (4\pi/g^2_{ab})$.

To translate this to the $SU(N) \to U(1)^{N-1}$ couplings, we remove the
trace piece to make the $(N-1)\times(N-1)$ matrix of low-energy
couplings
\be
\tau_0^{jk} = \tau^{jk} - \tau^{jN} - \tau^{Nk} + \tau^{NN}, \qquad
j,k=1,\ldots, N-1,
\ee
where $\tau$'s on the right refer to the $N\times N$ matrix of
couplings computed above in the $U(N)$ theory.  This subtraction just
reflects the fact that in the $SU(N)$ theory, one of the $N$ $U(1)$
subgroups is not independent of the others since their generators sum
to zero.  To include the theta angles (making $\tau$ complex), we can
simply drop the absolute value signs in (\ref{threshii}).  Although
there is not a unique way of assigning phases to the arguments of the
logarithms, the ambiguity is easily seen to be equivalent to an
electric-magnetic duality transformation (an integer shift of the low
energy $\tau^{jk}$'s).  Thus, the one-loop threshold contribution is
\bea\label{threshiii}
i\pi \tau^{jk}_0 
&=& \delta^{jk} \left( N \ln(\sqrt2 v_j) -
\sum_{a\neq j}^N \ln(v_a-v_j) \right) +
(1-\delta^{jk}) \ln(v_j-v_k) \nonumber\\ 
&& {} -\ln(v_j-v_N) -\ln(v_k-v_N) + N \ln(\sqrt2 v_N) 
- \sum_{\ell=1}^{N-1} \ln(v_\ell-v_N) .
\eea
Finally, this expression is not necessarily symmetric under interchange of
$j$ and $k$.  But in the low-energy effective action it multiplies
a term symmetric on these indices, so we should simply symmetrize
(\ref{threshiii}).

At the special vacuum, where $v_a = \omega^a$, (\ref{threshiii}) gives
\be\label{threshsv}
\tau^{jk}_0 = (1+\delta^{jk}) \left[{1\over2}
+ {i\over\pi} \ln (2N2^{-N/2}) \right]  + {i\over\pi} L^{jk},
\ee
where we have used the fact that $\prod_{k=1}^{N-1}(1-\omega^k) = N$
(see Appendix B), have made some integral electric-magnetic duality
shifts as in (\ref{emshift}), and where $L^{jk}$ is given in
(\ref{Ljkdef}).\footnote{The one-loop contribution to $\tau^{jk}$ in
the special vacuum was reported in \cite{mn9601} to be
$\tau^{jk}_0 = (1-\delta^{jk}) {1\over2} + (1+\delta^{jk})
{i\over\pi} \ln 2N + {i\over\pi} L^{jk}$.  After the $Sp(2N-2,\bz)$
electric-magnetic duality shift $\tau^{jk}_0 \to \tau^{jk}_0 +
\delta^{jk}$, this matches our result except for a $(1+\delta^{jk}) N
\ln 2 /(2\pi i)$ term.  This is due to the factor of $\sqrt2$ in the
hypermultiplet masses (\ref{hM}), which was neglected in
\cite{mn9601}.}  The matching between (\ref{threshsv}) and the
one-loop threshold contribution from the curve, given in 
Eq.~(\ref{taumatchcrv}) is achieved by choosing
\be
C_0=2^{N+2}.
\ee
Note that this value of $C_0$ differs from the one implicit in the
curve found in \cite{aps9505}, which gave $C_0=4$, whose normalization
is expected to satisfy our perturbative matching requirements (in
light of the result of \cite{fp9503} of no thresholds in the \drbar\
scheme).  The difference of a factor of ${\sqrt2}^{2N}$ can be traced
back to an incorrect extra factor of $\sqrt2$ in the superpotential 
mass term (\ref{massdef}) of \cite{aps9505}.

\subsection{One-instanton contribution}

The low energy ${\cal N}=2$ effective action is usefully expressed in
terms of the prepotential \cite{gsw78}, a scalar function of the
vector superfields, in terms of which the matrix of low energy
couplings is given by
\be
\tau^{jk} \equiv {1\over2} {\partial^{2}{\cal F} \over 
\partial V_j \partial V_k} ,
\ee
where the $V_j$ are a set of $N-1$ independent vevs on the Coulomb branch.
When the superpotential is given as a function of the $v_a$'s, as
in the instanton calculation of \cite{kms9804} to which we are
comparing the curve results, we can take the $N-1$ independent Coulomb
branch vevs to be the first $N-1$ of the $v_a$ used in
(\ref{tracev}).  The tracelessness constraint on the $v_a$ then implies
that $v_N$ is not an independent variable, and so the low-energy coupling
must be computed by
\be\label{taufromf}
\tau^{jk}={1 \over 2}
\left({\partial \over \partial v_j} 
- {\partial \over \partial v_N}\right)
\left({\partial \over \partial v_k} 
- {\partial \over \partial v_N}\right)
{\cal F}
\ee
where the partial derivatives are taken treating all $N$ $v_a$ as
independent variables.  The prepotential expanded about $q=0$
yields a classical piece, a one-loop correction and
instanton corrections:
\be
{\cal F} = {\ln q\over2\pi i} \left(\sum_{a=1}^N v_a^2\right)
+ \sum_{n=0}^{\infty} q^n {\cal F}_n .
\ee
Taking derivatives of this expansion as in (\ref{taufromf})
yields the instanton expansion (\ref{tauexp}) of the low energy
coupling $\tau^{jk}$.

The one-instanton prepotential for $N_f \leq 2N$ was calculated
in \cite{kms9804}.  Applying their result to $N_f = 2N$ with
massless matter gives the one-instanton prepotential:
\be\label{supot}
{\cal F}_1 = {C'_1 \pi^{2N-1}\over i2^{2N+2}} \sum_{c\neq d}
{(v_c+v_d)^{2N}\over (v_c-v_d)^2} \prod_{b\neq c,d}
{1\over (v_b-v_c)(v_b-v_d)}
\ee
where $b,c,d=1,\ldots,N$ and $C'_1$ is a scheme-dependent constant.
Differentiating this expression twice as in (\ref{taufromf}) and 
evaluating the resulting sums and products in the special vacuum,
we obtain after a lengthy computation (see Appendix B) the 
one-instanton contribution to $\tau^{jk}$ in the special vacuum:
\bea\label{uglysum}
\tau^{jk}_1 &=& {C'_1\pi^{2N}\over4\pi iN^2}
\Biggl\{ (1+\delta^{jk}) \left[ {N^2-1\over6} - {2N\choose N+1} 
{N^2 \over 2^{2N}} \right] \nonumber\\
&&\qquad {} - { 2\omega^j \over (1-\omega^j)^2 } 
- { 2\omega^k \over (1-\omega^k)^2 } 
+ (1-\delta^{jk}) { 2\omega^{j+k} \over (\omega^j - \omega^k)^2 }
\Biggr\} .
\eea
Comparing this with the one-instanton component of the exact
result, given in (\ref{taumatchcrvii}), implies
\bea
C'_1 &=& 2^{N+2}\pi^{-2N} ,\nonumber\\
C_1 &=& -8 \left[ 2^{2N} + {2N\choose N+1} \right] .
\eea

Thus, we have successfully matched the curve $\ttau^{jk}$ with the
semiclassical $\tau^{jk}$ in the special vacuum.  This determines the
first two coefficients $C_0$ and $C_1$ in the expansion of $\tq$ in
terms of $q$ of (\ref{qmatch}).  Note that this is a non-trivial matching
since the matching of the $N\times N$ matrix of low-energy effective
couplings forms a system of ${1\over2}N(N+1)$ equations in $2$
unknowns for each $N$. The existence of a solution for $C_0$ and $C_1$
for all $N$ thus provides an infinite number of physical checks
matching the predictions of the exact results and the semi-classical
results at the special vacuum.

\section*{Acknowledgments}
It is a pleasure to thank A. Buchel, V. Khoze, D. Pirjol, M. Mattis,
M. Strassler, and P. Yi for valuable discussions and comments.  This
work is supported in part by NSF grants PHY94-07194 and PHY95-13717.
The work of PCA and SP is supported in part by an A.P. Sloan
Foundation fellowship and by NSERC of Canada, respectively.

\section*{Appendix A: Details of curve computation of $\ttau^{jk}$}

Using the $\bz_N$ symmetry of the special vacuum, the computation of
$A^{\ell}_j$ and $B^{\ell k}$ can be reduced to integrals over the
paths $a$ and $b$ shown in Figure~1.  In particular, defining
\be
a_\ell \equiv \int_a \omega^{(\ell)},\qquad
b_\ell \equiv \int_b \omega^{(\ell)},
\ee
one finds that
\bea
A_j^\ell &=& 2 a_\ell \omega^{(j-1)\ell} , \nonumber\\
B^{\ell k} &=& 2 \left[ b_\ell \left(\omega^{(k-1)\ell}
-\omega^{(N-1)\ell}\right) 
+ a_\ell{\omega^{k\ell}-1\over\omega^\ell-1} \right] .
\eea
This implies that
\be
(A^{-1})^j_\ell = {1\over a_\ell} {\omega^{(1-j)\ell}-\omega^\ell
\over 2N},
\ee
so that
\bea\label{nothertau}
\ttau^{jk} &=& (A^{-1})^j_\ell B^{\ell k}
= \sum_{\ell=1}^{N-1} {\omega^{-j\ell}-1\over N}
\left[ {b_\ell \over a_\ell} (\omega^{k\ell}-1) + 
{\omega^{k\ell}-1\over 1-\omega^{-\ell}} \right],
\nonumber\\
&=& \Theta^{jk} + {1\over N} \sum_\ell {b_\ell \over a_\ell} 
(\omega^{-j\ell}-1) (\omega^{k\ell}-1) ,
\eea
where
\be
\Theta^{jk} = \left\{ \matrix{
1 & \mbox{for} & j\le k, \cr
0 & \mbox{for} & j >  k. \cr} \right.
\ee
The second line in (\ref{nothertau}) is obtained using the
identity
\be\label{identA}
\sum_{\ell\neq N} {\omega^{(a+1)\ell}\over \omega^\ell -1}
= {N-1\over2} -a + N \left[a\over N\right] ,
\ee
where $[x]$ denotes the integer part of $x$.  This identity
can be shown by writing the left side as
\be
\sum_{\ell\neq N} \left( {\omega^{(a+1)\ell}-1 \over 
\omega^\ell -1} + {1\over \omega^\ell -1} \right)
= \sum_{\ell\neq N} {1\over \omega^\ell -1} +
\sum_{\ell\neq N} \sum_{p=0}^a \omega^{p\ell} ,
\ee
and using the identity (\ref{lemmaii}) from Appendix B.

So we need to expand $a_\ell$ and $b_\ell$ about $\tq=0$.
More explicitly,
\bea
a_\ell &=& {1\over 2} \oint _{\widetilde\alpha_1}
{x^{\ell-1}dx \over \sqrt{(x^N-1)^2-\tq x^{2N}}} ,\nonumber\\
b_\ell &=& \int_0^{(1+\sqrt\tq)^{-1/N}}
{x^{\ell-1}dx \over \sqrt{(x^N-1)^2-\tq x^{2N}}} .
\eea
$a_\ell$ can be evaluated using the residue theorem, with the 
result:
\be
a_\ell={i\pi\over N}\left[1+ {\ell(N+\ell)\over 4N^2} \tq 
+{\cal O}(\tq^2)\right] .
\ee
Since $b_\ell$ is singular at $\tq=0$, it must be 
evaluated by a less direct method.  First, we recognize $b_\ell$ as
the hypergeometric function \cite{gr65}
\be
b_\ell ={1 \over N(1+\sqrt{\tq})^{\ell/N}} \,
B\left({\ell\over N},{1\over2}\right)\,\cdot\, 
F\left[{1\over2},{\ell\over N},{\ell\over N}+{1\over2},
{1-\sqrt{\tq}\over 1+\sqrt{\tq}}\right] .
\ee
Expanding about $\tq=0$ gives
\bea
b_\ell &=& {1\over N} \left(1-{\ell\over N} \sqrt{\tq} 
+{\ell(\ell+N) \over 2N^2} \tq + \ldots \right) \nonumber\\
&& {}\times \Biggl\{ 
\left[-2\gamma-\psi\left({1\over2}\right) 
-\psi\left({\ell\over N}\right) - \ln{y}\right]
\left(1+{\ell\over2N} y + {3\ell(\ell+N)\over16N^2}y^2 
+\ldots \right) \nonumber\\ 
&& \qquad {} - {1\over2} y 
+ {\ell^2-5\ell N-3N^2\over 16N^2} y^2+\ldots\Biggr\} ,
\eea 
where
\be
y \equiv {2\sqrt\tq \over 1+\sqrt\tq},
\ee
$\gamma=.577215...$ is the Euler constant, and $\psi(x) = 
d\ln\Gamma(x)/dx$.  Then, to order $\tq$ we have, using
$\gamma + \psi(1/2) = -2\ln2$,
\be
{b_\ell \over a_\ell} = {\ln\tq\over2\pi i} + {i\over\pi}
\left(\ln2-\gamma-\psi(\ell/N)\right) + {i\over\pi}
\left({\ell^2-\ell N - N^2 \over 4N^2}\right)\tq .
\ee
Inserting this in (\ref{nothertau}), using Gauss' theorem
for $\psi$ of a rational argument,
\be
\psi\left(\ell\over N\right) +\gamma = -\ln N -{i\pi\over2} 
{\omega^\ell+1\over\omega^\ell-1}+ {1\over2} \sum_{j=1}^{N-1}
\left(\omega^{j\ell}+\omega^{-j\ell}\right)\,
\ln\left(\omega^{j/2} - \omega^{-j/2} \right),
\ee
and doing the sum then gives (\ref{tausvcurve}).

\section*{Appendix B: Details of the one-instanton contribution 
to $\tau^{jk}$}

The one-instanton prepotential (\ref{supot}) can be rewritten as
\be\label{supoti}
{\cal F}_1 = {\pi^{2N} C'_1 \over 4\pi i} 
\left( F_1 - 2^{-2N} F_2 \right)
\ee
with
\bea\label{supotii}
F_1 &\equiv& \sum_{c=1}^N \left( {v_c^N \over \Pi_c} \right)^2 ,
\nonumber\\
F_2 &\equiv& \sum_{c,d=1}^N {1\over\Pi_c\Pi_d} (v_c+v_d)^{2N} , 
\eea
and
\be
\Pi_c \equiv \prod_{b\neq c} (v_b-v_c) .
\ee

$\tau_1^{jk}$ is obtained from ${\cal F}_1$ by taking second partial
derivatives as discussed in (\ref{taufromf}),
\be\label{nothertauf}
\tau^{jk}={1\over2}
( \partial_j - \partial_N ) ( \partial_k - \partial_N ) {\cal F}_1 ,
\ee
where we have introduced the notation $\partial_a = \partial/\partial
v_a$.  We therefore need to compute $\partial_a \partial_b F_{1,2}$ at
the special vacuum ($v_a = \omega^a$ where $\omega=e^{2\pi i/N}$).

We start with $F_1$.  Taking derivatives of (\ref{supotii}) and
evaluating at the special vacuum gives
\bea
&&\partial_a\partial_b F_1 = \nonumber\\
&& \sum_c{1\over\Pi_c^2} \Biggl\{
\left({2N\delta_{ac} \over \omega^c} 
- {2(1-\delta_{ac})\over \omega^a-\omega^c}
+ \sum_{d\neq c} {2\delta_{ac}\over \omega^d-\omega^c}\right)
\left({2N\delta_{bc} \over \omega^c} 
- {2(1-\delta_{bc})\over \omega^b-\omega^c}
+ \sum_{d\neq c} {2\delta_{bc}\over \omega^d-\omega^c}\right)
\nonumber\\
&&\qquad\qquad{}-{2N\delta_{ac} \delta_{bc}\over \omega^2_c} 
+{2(1-\delta_{ac})(\delta_{ab}-\delta_{cb})\over 
(\omega^a-\omega^c)^2}
-\sum_{d\neq c} {2\delta_{ac}(\delta_{db}-\delta_{cb})
\over (\omega^d-\omega^c)^2}
\Biggr\} .
\eea
The following lemmas prove to be useful in the evaluation of the above
sums:
\be\label{lemmai}
\Pi_c = (-)^{N-1}{N \over \omega^c},
\ee
\be\label{lemmaii}
\sum _{f \neq c}{1 \over \omega^f -\omega^c} 
= {1-N \over 2 \omega^c} 
\ee
and
\be\label{lemmaiii}
\sum_{f \neq c}{1 \over (\omega^f -\omega^c)^2} 
= {(N-1)(5-N) \over 12\omega^{2c}},
\ee
where (\ref{lemmai}) is evaluated in the special vacuum.
(\ref{lemmai}) follows from the fact that $\prod_{a\neq N}(z-\omega^a)
= (z^N-1)/(z-1)$ and $\lim_{z\to1} (z^N-1)/(z-1) = N$.  Similarly,
(\ref{lemmaii}) follows from $\sum_{f\neq N} (1-\omega^f)^{-1} =
N^{-1} \lim_{z\to1} \partial_z[(z^N-1)/(z-1)] = (N-1)/2$.  Finally,
(\ref{lemmaiii}) follows from $[\sum_{f\neq N}(1-\omega^f)^{-1}]^2-
\sum_{f\neq N}(1-\omega^f)^{-2} = N^{-1} \lim_{z\to1} \partial^2_z
[(z^N-1)/(z-1)]$.  Then, somewhat lengthy algebra gives
\be\label{fi}
\partial_a  \partial_b F_1 = {1\over N^2} \left[
2(2N-1) + \delta_{ab} {1\over3} (N^2-1) + (1-\delta_{ab}) 
{4\omega^a \omega^b \over (\omega^a - \omega^b)^2}\right] .
\ee

The derivatives of $F_2$ are more complicated:
\be\label{effii}
\partial_a\partial_b F_2 = 2 \sum_{L=0}^{2N}
{2N\choose L} \left[ f_{2N-L} \partial_a\partial_b f
+ \partial_a f_{2N-L} \cdot \partial_b f \right] ,
\ee
where
\be
f_L \equiv \sum_c {v_c^L \over \Pi_c} .
\ee
Using our lemmas (\ref{lemmai}--\ref{lemmaiii}) and the identity 
(\ref{identA}) from Appendix A, after some algebra one finds in the 
special vacuum that
\be\label{identB}
f_L = (-)^{N-1} \left\{ 
\matrix{1 & L=-1\ \mbox{mod}\ N,\cr 0 &\mbox{otherwise},\cr}
\right.
\ee
\be
\partial_a f_L = (-)^{N-1} \omega^{aL} \left\{ 
\matrix{0 & 0\le L< N,\cr 1 & N\le L< 2N,\cr 
2 & L= 2N,\cr} \right.
\ee
and
\be\label{interiii}
\partial_a \partial_b f_L = (-)^{N-1} {L-1\over N} 
\left( 1 + \delta_{ab}(L-N) \right)
\qquad\mbox{if}\quad L=1\ \mbox{mod}\ N.
\ee
{}From (\ref{effii}) and (\ref{identB}) we only need the $L=1$ mod $N$ 
case in (\ref{interiii}).  Inserting these into (\ref{effii}) gives
\be\label{fii}
\partial_a \partial_b F_2 =  
2{2N \choose N+1}\left[{2N+1 \over N} +\delta_{ab} \right] .
\ee
Finally, substituting (\ref{fi},\ref{fii}) into (\ref{nothertauf}) gives 
(\ref{uglysum}).


\begin{thebibliography}{99}

\bibitem{sw9407}N. Seiberg and E. Witten, hep-th/9407087, \NP{\bf
B426} (1994) 19.
% N=2 SU(2) YM curve

\bibitem{sw9408}N. Seiberg and E. Witten, hep-th/9408099, \NP{\bf
B431} (1994) 484.
% N=2 and N=4 SU(2) SI curve

\bibitem{ahsw9607}H. Aoyama, T. Harano, M. Sato and S. Wada,
hep-th/9607076, \PL{\bf B388} (1996) 331.
% N=2 SU(2) AF & SI shifts

\bibitem{dkm9607}N. Dorey, V.V. Khoze and M.P. Mattis, hep-th/9607202,
\PR{\bf D54} (1996) 7832.
% N=2 SU(n) SI (prob)

\bibitem{hs9608}T. Harano and M. Sato,
hep-th/9608060, \NP{\bf B484} (1997) 167.
% referee's suggestion

\bibitem{is9609}K. Ito and N. Sasakura, hep-th/9609104, \MPL{\bf A12}
(1997) 205.
% N=2 SU(3) AF & SI shifts

\bibitem{dkm9612}N. Dorey, V.V. Khoze and M.P. Mattis, hep-th/9612231,
\PL{\bf B396} (1997) 141.
% N=4 mass-deformed (prob)

\bibitem{kms9804}V.V. Khoze, M.P. Mattis and M.J. Slater,
hep-th/9804009, \NP{\bf B536} (1998) 69.
% N=2 SU(n) SI (prob)

\bibitem{as9509}P.C. Argyres and A.D. Shapere, hep-th/9509175, \NP{\bf
B461} (1996) 437.
% match mn9507 to aps9505

\bibitem{mn9601}J.A. Minahan and D. Nemeschansky, hep-th/9601059,
\NP{\bf B468} (1996) 72.
% match mn9507 to aps9505; threshold corr at sv

\bibitem{dkm9611}N. Dorey, V.V. Khoze and M.P. Mattis, hep-th/9611016,
\NP{\bf B492} (1997) 607.
% N=2 SU(2) SI shift

\bibitem{s9701}M.J. Slater, hep-th/9701170, \PL{\bf B403} (1997) 57.
% N=2 SU(n) AF shifts

\bibitem{aps9505}P.C. Argyres, M.R. Plesser and A.D. Shapere,
hep-th/9505100, \PRL{\bf 75} (1995) 1699.
% N=2 SU(n) SI curve

\bibitem{ho9505}A. Hanany and Y. Oz, hep-th/9505075, \NP{\bf B452}
(1995) 283
% alternative N=2 SU(n) SI curve

\bibitem{mn9507}J.A. Minahan and D. Nemeschansky, hep-th/9507032,
\NP{\bf B464} (1996) 3.
% alternative N=2 SU(3) SI curve

\bibitem{s79}W. Siegel, \PL{\bf B84} (1979) 193; D.M. Capper,
D.R.T. Jones and P. van Nieuwenhuizen, \NP{\bf B167} (1980) 479.
% DRbar scheme

\bibitem{h86}G. 't Hooft, \PRe{\bf 142} (1986) 357; S. Cordes, \NP{\bf
B273} (1986) 629.
% DRbar instanton calcs

\bibitem{fp9503}D. Finnell and P. Pouliot, hep-th/9503115, \NP{\bf
B453} (1995) 225.
% DRbar instanton calcs

\bibitem{a9706}P.C. Argyres, hep-th/9706095, \ATMP{\bf 2} (1998) 293; 
P.C. Argyres and A. Buchel, hep-th/9910125.
% meaning of Sduality

\bibitem{ay9601}O. Aharony and S. Yankielowicz, hep-th/9601011, 
\NP{\bf B743} (1996) 93. 
% no ``classical'' coupling point

\bibitem{w80}S. Weinberg, \PL{\bf B91} (1980) 51; L. Hall, \NP{\bf
B178} (1981) 75.
% threshold corrections

\bibitem{mv9308}S.P. Martin and M.T. Vaughn, hep-ph/9308222, \PL{\bf
B318} (1993) 331.
% susy threshold corrections

\bibitem{gsw78}R. Grimm, M. Sohnius and J. Wess, \NP{\bf 133} (1978)
275.
% N=2 superspace

\bibitem{gr65}I.S. Gradshteyn and I.M. Ryzhik, {\it Table of Integrals, 
Series, and Products}, Academic Press, 1965.

\end{thebibliography}
\end{document}